\documentclass[12pt]{article}
\pdfoutput=1

\usepackage{amsmath}
\usepackage{amsfonts}
\usepackage{amscd}
\usepackage{amsthm}
\usepackage{setspace}

\usepackage{graphicx}
\usepackage{authblk}
\usepackage{caption}
\usepackage{ytableau}
\usepackage{mathtools}

\def\Z{\mathbb{Z}}

\def\C{\mathbb{C}}
\def\P{\mathbb{P}}

\begin{document}

\begin{titlepage}

\begin{flushright}
KEK-TH-2195
\end{flushright}

\vskip 1cm

\begin{center}

{\bf \Large Extremal 1/2 Calabi--Yau 3-folds and six-dimensional F-theory applications}

\vskip 1.2cm

Yusuke Kimura$^1$ 
\vskip 0.6cm
{\it $^1$KEK Theory Center, Institute of Particle and Nuclear Studies, KEK, \\ 1-1 Oho, Tsukuba, Ibaraki 305-0801, Japan}
\vskip 0.4cm
E-mail: kimurayu@post.kek.jp

\vskip 2cm
\abstract{We discuss a method for classifying the singularity types of 1/2 Calabi--Yau 3-folds, a family of rational elliptic 3-folds introduced in a previous study in relation to various U(1) factors in 6D F-theory models. A projective dual pair of del Pezzo manifolds recently studied by Mukai is used to analyze the singularity types. In particular, we studied the maximal rank seven singularity types of 1/2 Calabi--Yau 3-folds. The structures of the singular fibers are analyzed using blow-ups. Double covers of the 1/2 Calabi--Yau 3-folds yield elliptic Calabi--Yau 3-folds and applications to six-dimensional $N = 1$ F-theory on the Calabi--Yau 3-folds are also discussed. The deduced singular fibers have applications in studying the gauge groups formed in 6D F-theory compactifications. The blow-up methods used to analyze the singular fibers and sections utilized in this research might have applications in studying the U(1) factors and hypermultiplets charged under U(1) in 6D F-theory.}  

\end{center}
\end{titlepage}

\tableofcontents
\section{Introduction}
U(1) symmetry is important in realizing the grand unified theory (GUT) because the presence of a U(1) symmetry helps explain a few of the characteristic properties of GUT, such as the mass hierarchies of the quarks and leptons, and a suppression of the proton decay. In F-theory \cite{Vaf, MV1, MV2}, information of U(1) gauge symmetry can be extracted from the geometry of the compactification space. F-theory is compactified on elliptic fibrations, and the modular parameter of the tori as fibers of the elliptic fibration is identified with axiodilaton, enabling the axiodilaton to exhibit $SL(2,\Z)$ monodromy. The $SL(2,\Z)$ symmetry possessed by type IIB superstrings is realized in a geometric manner in the formulation of F-theory. 
\par A certain structure of elliptic fibration, a global section, relates directly to the U(1) gauge symmetry that forms in F-theory. When one can choose a point in every elliptic fiber of the fibration, and the chosen point can be moved throughout the base space of the fibration, a genus-one fibration having such a structure is said to admit a global section, yielding a copy of the base space inside the total space of the fibration. When an elliptic fibration has a global section, the set of global sections forms a group, which is known as the ``Mordell--Weil group.'' The rank of the Mordell--Weil group of an elliptic fibration, in the context of physics, is related to the U(1) gauge group; the rank yields the number of U(1) factors formed in F-theory on that elliptic fibration \cite{MV2}. 
\par F-theory models on elliptic fibrations having a global section have been intensively studied, e.g., in \cite{MorrisonPark, MPW, BGK, BMPWsection, CKP, BGK1306, CGKP, CKP1307, CKPS, Mizoguchi1403, AL, EKY1410, LSW, CKPT, CGKPS, MP2, MPT1610, BMW2017, CL2017, BMW1706, KimuraMizoguchi, Kimura1802, LRW2018, TasilectWeigand, MizTani2018, TasilectCL, Kimura1810, CMPV1811, TT2019, Kimura1902, Kimura1903, EJ1905, LW1905, Kimura1910, CKS1910, Kimura1911, AFHRZ2001}. U(1) gauge symmetry \footnote{Recent studies of F-theory models in which one or more factors of U(1) form can be found, for example, in \cite{MorrisonPark, BMPWsection, CKP, CGKP, BMPW2013, CKPS, MTsection, MT2014, KMOPR, BGKintfiber, CKPT, GKK, MPT1610, LS2017, Kimura1802, TT2018, CLLO, CMPV1811, TT2019, Kimura1908, Kimura1910, Kimura1911, OS2019, AFHRZ2001}.} has also been investigated in F-theory. 
\par In this study, we mainly focus on six-dimensional (6D) F-theory. 

\vspace{5mm}

\par Elliptic Calabi--Yau 3-folds of various Mordell--Weil ranks have recently been constructed by taking double covers of certain class of rational elliptic 3-folds, which are referred to as ``1/2 Calabi--Yau 3-folds'' \cite{Kimura1910}. F-theory on this type of Calabi--Yau 3-folds yields 6D $N = 1$ theories with various numbers of U(1) factors \cite{Kimura1910}. A general construction scheme of elliptically fibered Calabi--Yau 3-folds by taking double covers of ``1/2 Calabi--Yau 3-folds'' is discussed in \cite{Kimura1910}, and some explicit examples of 1/2 Calabi--Yau 3-folds with specific singularity types and Calabi--Yau 3-folds obtained as their double covers are discussed in \cite{Kimura1910}.
\par The aim of this study is to discuss a strategy to classify the singularity types of the elliptic Calabi--Yau 3-folds constructed as double covers of 1/2 Calabi--Yau 3-folds. This result translates, in string theoretic language, to non-Abelian gauge groups \footnote{Discussion of the correspondence between the non-Abelian gauge groups forming on the 7-branes in F-theory on an elliptic fibration and the fiber types can be found in \cite{MV2, BIKMSV}.} forming in 6D F-theory compactifications on them. Furthermore, this analysis yields a description of the singular fibers \footnote{Kodaira classified the types of the singular fibers of the elliptic surfaces in \cite{Kod1, Kod2}. The authors of \cite{Ner, Tate} discussed methods to determine the types of the singular fibers of the elliptic surfaces.} and sections of 1/2 Calabi--Yau 3-folds and Calabi--Yau 3-folds as double covers. Therefore, the analysis also directly relates to U(1) gauge groups and matter fields arising in 6D F-theory compactifications. 
\par There are some obstacles to studying the singularities of 1/2 Calabi--Yau 3-folds. The 1/2 Calabi--Yau 3-folds were constructed as blow-ups of $\P^3$ at the intersection points of three quadrics \cite{Kimura1910}. When studying the singular fibers of the resulting 1/2 Calabi--Yau 3-folds without resolving the singularity, only two conics meeting in two points can be found, and this generally does not determine the types of singular fibers. Determining the types of singular fibers given the equations of three quadrics requires multiple stages of resolutions, which makes an analysis of the singular fibers difficult to achieve. 
\par To resolve this difficulty in analyzing the singularity types of 1/2 Calabi--Yau 3-folds directly, we relate the problem of classifying the singularity types of the 1/2 Calabi--Yau 3-folds to those of quartic curves in $\P^2$ by considering the ``projective duals.'' An interesting mathematical result observed by Mukai in \cite{Muk, Mukai2008, Mukai2019}, when applied to the 1/2 Calabi--Yau 3-folds, reveals that the classification of the 1/2 Calabi--Yau 3-folds is actually equivalent to the classification of the singularity types of the quartic curves in $\P^2$. The singularity types of the 1/2 Calabi--Yau 3-folds and those of plane quartic curves are actually equivalent based on the notion of the ``projective duality'' \cite{Mukai2019}. Making use of this duality, the classification problem of the singularity types of 1/2 Calabi--Yau 3-folds, which appeared to be obscure and somewhat difficult, begins to become more transparent, enabling us to find a way to resolve the problem. By making use of this method, we classify the singularity types of the 1/2 Calabi--Yau 3-folds when the rank is maximal, and furthermore, through this approach the singular fibers are described in detail. Based on this analysis, the structures of the global sections are also described. The discussion of these are the main goals of this paper. 
\par Because a singularity type of Calabi--Yau 3-fold as a double cover of the original 1/2 Calabi--Yau 3-fold is identical to the singularity type of the original Calabi--Yau 3-fold \cite{Kimura1910}, the classification results of singularity type 1/2 Calabi--Yau 3-folds also yield a classification of the singularity types of the Calabi--Yau 3-folds as double covers. From the viewpoint of F-theory, these results yield the non-Abelian gauge groups forming on the 7-branes in the 6D F-theory on the Calabi--Yau 3-folds. When the classification scheme is applied to 1/2 Calabi--Yau 3-folds of singularity ranks of strictly lower than seven, by taking their double covers, 6D $N = 1$ F-theory models with (multiple) U(1) factors can also be analyzed. Because our analysis here includes a description of the singular fibers, the analysis can also be applied to the matter spectra in the 6D compactifications. 
\par Obtaining the Weierstrass equations of the 1/2 Calabi--Yau 3-folds can be considerably difficult as pointed out in \cite{Kimura1910}. Although obtaining the Weierstrass equations is useful in determining the gauge groups and the matter spectra, we take an approach to deduce the singularity types and the singular fibers directly from the defining equations of the three quadrics, which specify the complex structure of the 1/2 Calabi--Yau 3-fold described in this paper. 
\par We focus on 1/2 Calabi--Yau 3-folds with a singularity rank of seven, which is the maximal rank for the 1/2 Calabi--Yau 3-folds \cite{Kimura1910}, and we classify the singularity types of such 1/2 Calabi--Yau 3-folds in this study. We refer to 1/2 Calabi--Yau 3-folds having the maximal rank-seven singularity types as ``extremal 1/2 Calabi--Yau 3-folds.'' The rank-seven singularity types of the quartic curves in $\P^2$ were realized in \cite{DolgachevAlgGeom} and the classification of the rank-seven singularities consists of six types \cite{DolgachevAlgGeom}. These six types yield the singularity types of the extremal 1/2 Calabi--Yau 3-folds via applying the method in \cite{Muk, Mukai2008, Mukai2019}. Our classification method also applies to 1/2 Calabi--Yau 3-folds having lower singularity ranks.
\par By analyzing the extremal 1/2 Calabi--Yau 3-folds, we demonstrate that the structures of the singular fibers and the global sections can be explicitly seen by conducting blow-ups. Among the classified singularity types of the extremal 1/2 Calabi--Yau 3-folds, we study in detail two singularity types to describe the singular fibers and sections. These two types require relatively shallow levels of blow-ups to understand the structures of the singular fibers. However, other singularity types require deeper levels of blow-ups, as mentioned in section \ref{sec3.3}, and our study suggests that through multiple stages of blow-ups the singular fibers can be described in manners similar to the two singularity types. 
\par Local model buildings \cite{DWmodel, BHV1, BHV2, DW} have been emphasized in recent studies on F-theory. The global aspects of the models, however, need to be studied to discuss issues of gravity and problems pertaining to the early universe. The global aspects of the compactification geometry are analyzed in this study. 
\par Studies on geometric structures of elliptically fibered 3-folds can be found in \cite{Nak, DG, G}.

\vspace{5mm}

\par This paper is structured as follows: We discuss a method for classifying the singularity types of 1/2 Calabi--Yau 3-folds in section \ref{sec2.1}. The maximal rank seven singularity types of 1/2 Calabi--Yau 3-folds are described in sections \ref{sec2.2} through \ref{sec2.7}. There are six types of such singularities. Two of these singularity types are analyzed in detail in sections \ref{sec3.1} and \ref{sec3.2}. We demonstrate that the singular fibers and sections can be described after eight blow-ups. We also mention the remaining rank seven singularity types in section \ref{sec3.3}. The F-theory application is discussed in section \ref{sec4}. Singularity types of 1/2 Calabi--Yau 3-folds of ranks lower than seven and U(1) factors in 6D F-theory are also mentioned. Concluding remarks and remaining problems are mentioned in section \ref{sec5}. Problems that are possibly related to the swampland conditions are also discussed. Reviews of recent progress of the swampland criteria can be found in \cite{BCV1711, Palti1903}. In \cite{Vafa05, AMNV06, OV06}, the authors discussed the notion of the swampland. A new consistency condition on 6D $N=1$ quantum gravity theories was recently discussed in \cite{KSV1905}. The authors of \cite{KT0910, KMT1008, PT1110, Taylor1104} discussed the possible combinations of distinct matter fields and gauge symmetries for quantum gravity theories in 6D with $N=1$ supersymmetry.  

\section{Singularity types of 1/2 Calabi--Yau 3-folds and projective duality}
\label{sec2}

\subsection{Method to deduce the equations of three quadrics of 1/2 Calabi--Yau 3-folds}
\label{sec2.1}
The 1/2 Calabi--Yau 3-folds constructed in \cite{Kimura1910} are rational elliptic 3-folds obtained by blowing up $\P^3$ at the intersection points of three quadrics. The base surface of a 1/2 Calabi--Yau 3-fold is isomorphic to $\P^2$, and taking the ratio of three quadrics $q_1, q_2, q_3$, $[q_1:q_2:q_3]$, yields the projection. Taking double covers of the 1/2 Calabi--Yau 3-folds (ramified along appropriate degree 8 hypersurfaces) yields elliptically fibered Calabi--Yau 3-folds, F-theory compactifications upon which 6D $N = 1$ theories are provided \cite{Kimura1910}. 
\par The singularity types of the original 1/2 Calabi--Yau 3-fold and the Calabi--Yau 3-fold as its double cover are identical \cite{Kimura1910}, thus determining the singular fibers and the singularity types of the 1/2 Calabi--Yau 3-folds also determines the singularity types of elliptic Calabi--Yau 3-folds as double covers. This in principle determines the non-Abelian gauge symmetries that arise in F-theory on the Calabi--Yau 3-folds \footnote{Whether the singular fibers are split, non-split, or semi-split \cite{BIKMSV} also needs to be specified to deduce the precise gauge group.}. 
\par Pairs of algebraic varieties of different dimensions that are projective duals were studied in \cite{Mukai2019}. Applying the analysis in \cite{Mukai2019} to 1/2 Calabi--Yau 3-folds, the classifications of the singularity types of 1/2 Calabi--Yau 3-folds and the singularities of the quartic curves in $\P^2$ are found to be identical. Making use of this mathematical observation reduces the classification problem of the singularity types of 1/2 Calabi--Yau 3-folds to that of the singularities of the quartic curves in $\P^2$. 
\par In \cite{Mukai2019}, Mukai studied several pairs of del Pezzo manifolds, which are projective duals to each other. We apply one of these pairs, $(X_2, Y_8)$, to 1/2 Calabi--Yau 3-folds. Here, $X_2$ is a del Pezzo manifold of dimension ${\rm dim}_{\C} X_2 =9$, and del Pezzo manifold $Y_8$ has dimensions of ${\rm dim}_{\C} Y_8=3$. $X_2$ is a double cover of $\P^9$ ramified over quartic hypersurface $D_4$, and $Y_8$ is the Veronese 3-fold $v_2(\P^3)$ embedded inside $\P^9$ \cite{Mukai2019}. Moreover, $X_2$ and $Y_8$ are projective duals to each other \cite{Mukai2019}. Here, $X_2$ cut out by seven hyperplanes yields a double cover of $\P^2$ ramified over a quartic curve, which is the degree-two del Pezzo surface \footnote{This surface is isomorphic to the blow-up of $\P^2$ at seven points of a general position.}. Because the projective dual of a hyperplane is a point, the operation of cutting $X_2$ by seven hyperplanes on the $Y_8$ side corresponds to the choice of seven points. The seven points span $\P^6$, and therefore cutting $X_2$ by seven hyperplanes corresponds on the $Y_8$ side to taking the intersection of $Y_8=v_2(\P^3)$ and $\P^6$ inside $\P^9$, $v_2(\P^3)\cap \P^6 \subset \P^9$. The intersection is equivalent to cutting $v_2(\P^3)$ by three hyperplanes inside $\P^9$. Because $v_2(\P^3)$ is a Veronese embedding of $\P^3$ into $\P^9$, taking the intersection $v_2(\P^3)\cap \P^6 \subset \P^9$ is equivalent to the intersection of the three quadrics in $\P^3$ \cite{Mukai2019}. $X_2$ cut out by seven hyperplanes is isomorphic to the degree-two del Pezzo surface \cite{Mukai2019} which is a double cover of $\P^2$ ramified over a quartic curve, and the blow-up of $\P^3$ at the intersection points of the three quadrics yields the Jacobian of the degree-two del Pezzo surface \cite{Muk, Mukai2008, Mukai2019}. Therefore, the singularity types of the quartic curves in $\P^2$ are identical to the singularity types of the 1/2 Calabi--Yau 3-folds, and the correspondence is manifest through the projective duality discussed in \cite{Mukai2019}. 
\par To describe the correspondence of the equations of three quadrics of a 1/2 Calabi--Yau 3-fold and the quartic curve in $\P^2$, when the determinantal representation of a plane quartic curve is given as a symmetric 4 $\times$ 4 matrix, matrix elements correspond to the equations of the three quadrics in $\P^3$ via the method discussed in \cite{Muk, Mukai2008, Mukai2019}. When one finds the determinantal representation of a quartic curve, equations of the three quadrics of the dual 1/2 Calabi--Yau 3-fold can be deduced from the determinantal representation.

\vspace{5mm}

\par The maximal singularity rank of 1/2 Calabi--Yau 3-folds is seven \cite{Kimura1910}. We classify the rank seven singularity types of the 1/2 Calabi--Yau 3-folds utilizing the method in \cite{Muk, Mukai2008, Mukai2019}. The 1/2 Calabi--Yau 3-folds possessing the maximal rank seven singularity types are referred to as extremal 1/2 Calabi--Yau 3-folds in this note. 
\par
\begin{figure}
\begin{center}
\includegraphics[height=10cm, bb=0 0 960 540]{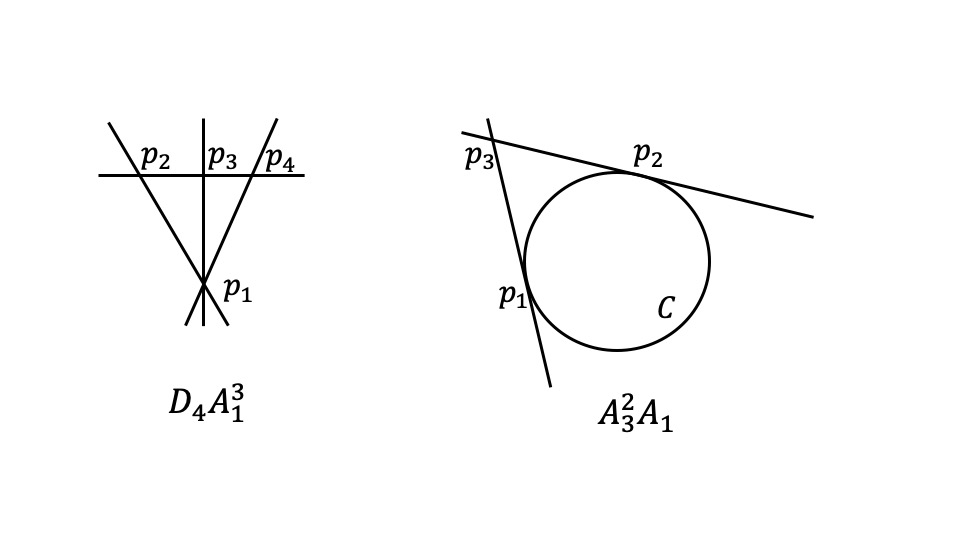}
\caption{\label{figD4sum3A1}The left image describes the quartic curve, which is reducible into three lines meeting at one point and a line. This curve has $D_4A_1^3$ singularity \cite{DolgachevAlgGeom}. The point at which three lines simultaneously meet yields a $D_4$ singularity, and each of the other three intersection points yields $A_1$ singularity. The right image describes a quartic curve reducible into one conic and two tangents. The quartic curve on the right has $A_3^2A_1$ singularity \cite{DolgachevAlgGeom}. Two points at which two lines are tangent to the conic yield two $A_3$ singularities. The intersection of two lines yields $A_1$ singularity.}
\end{center}
\end{figure}

\begin{figure}
\begin{center}
\includegraphics[height=10cm, bb=0 0 960 540]{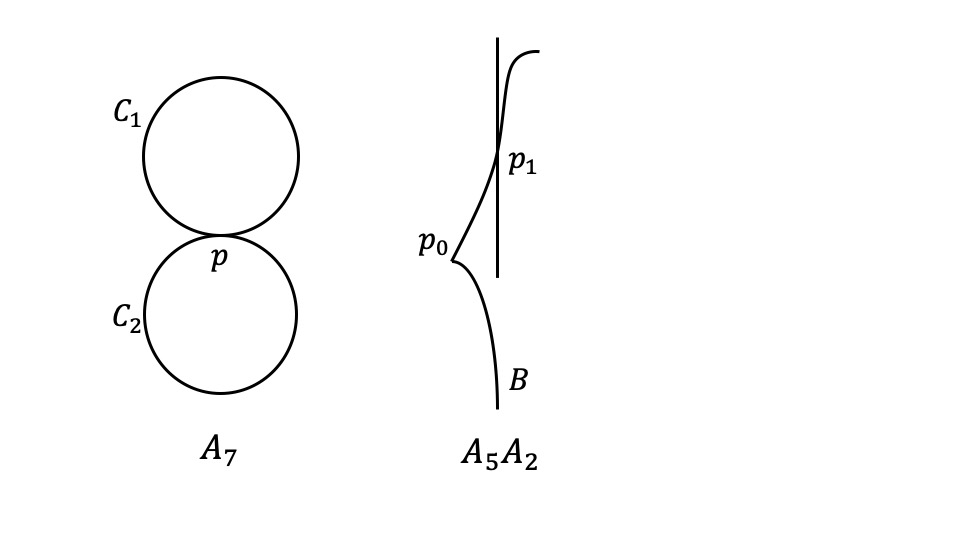}
\caption{\label{figA7}The left image describes a quartic curve reducible into two conics meeting at a point. The left quartic curve has $A_7$ singularity \cite{DolgachevAlgGeom}. The intersection point yields $A_7$ singularity. The right image describes a cubic with a cusp and a line tangent to the flex. The quartic curve on the right has $A_5A_2$ singularity \cite{DolgachevAlgGeom}. A cusp yields $A_2$ singularity, and the tangent point of the flex and the line yields $A_5$ singularity.}
\end{center}
\end{figure}

\begin{figure}
\begin{center}
\includegraphics[height=10cm, bb=0 0 960 540]{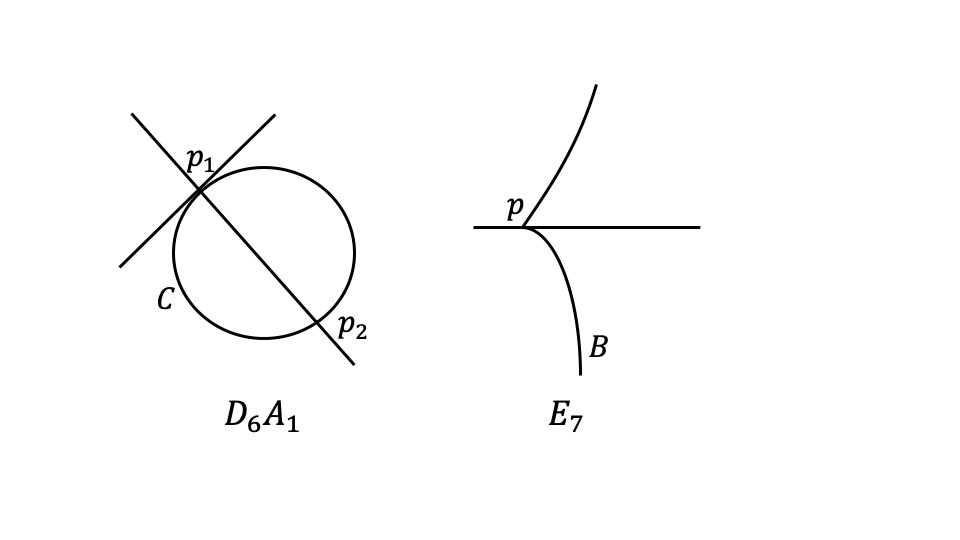}
\caption{\label{figD6A1}The left image describes a quartic curve reducible into a conic, a line tangent to the conic, and another line through the tangent point. The left quartic curve has $D_6A_1$ singularity \cite{DolgachevAlgGeom}. The right image is a cubic with a cusp and cuspidal tangent. The right quartic curve has $E_7$ singularity \cite{DolgachevAlgGeom}.}
\end{center}
\end{figure}

 The classification of the singularity types of the quartic curves in $\P^2$ can be found in \cite{DolgachevAlgGeom}. Among the singularities, seven is the maximal rank and there are six rank-seven singularity types \cite{DolgachevAlgGeom}: $D_4A_1^3$, $A_3^2A_1$, $A_5A_2$, $A_7$, $D_6A_1$, $E_7$. Therefore, from the argument given, we found that the extremal 1/2 Calabi--Yau 3-folds have six singularity types via applying the method in \cite{Muk, Mukai2008, Mukai2019}. The corresponding six singularity types of the plane quartics are shown in Figures \ref{figD4sum3A1}, \ref{figA7}, and \ref{figD6A1}. 

\vspace{5mm}
   
\par As presented in Figures \ref{figD4sum3A1}, \ref{figA7}, and \ref{figD6A1}, the quartic curves with rank seven singularities are reducible into a cubic and a line, two conics, lines and a conic, or four lines. Considering their determinantal representations, these situations correspond to a 4 $\times$ 4 matrix reducible into smaller blocks of submatrices.  
\par Given the series of observations, we are now ready to deduce the equations of the three quadrics of the extremal 1/2 Calabi--Yau 3-folds, which are described in sections \ref{sec2.2}, \ref{sec2.3}, \ref{sec2.4}, \ref{sec2.5}, \ref{sec2.6}, and \ref{sec2.7}. 

\subsection{$D_4A_1^3$ singularity}
\label{sec2.2}
\par We determine the equations of the three quadrics yielding the extremal 1/2 Calabi--Yau 3-fold with the singularity type $D_4A_1^3$, when $\P^3$ is blown up at the intersections of the three quadrics. 
\par The dual quartic curve with the $D_4A_1^3$ singularity is the sum of three lines meeting at a point and a line \cite{DolgachevAlgGeom}, as presented in Figure \ref{figD4sum3A1}. This quartic curve is given by the following equation:
\begin{equation}
\label{quartic curve in 2.2}
\lambda(\lambda-\mu)\mu\nu=0,
\end{equation}
where $[\lambda:\mu:\nu]$ denotes the coordinates of $\P^2$. The three lines $\lambda=0$, $\mu=0$, and $\lambda-\mu=0$ yield three lines meeting at a point, which we denote as $p_1$. The determinantal representation of the quartic curve (\ref{quartic curve in 2.2}) is given as follows:
\begin{equation}
\label{detrep in 2.2}
\begin{pmatrix}
\lambda & 0 & 0 & 0 \\
0 & \mu-\lambda & 0 & 0 \\
0 & 0& -\mu & 0 \\
0 & 0& 0 & \nu
\end{pmatrix}.
\end{equation}
\par The equations of the three quadrics, $q_1, q_2, q_3$, can be deduced from the determinantal representation (\ref{detrep in 2.2}). The entries of the matrix (\ref{detrep in 2.2}) where $\lambda$ appears yield the coefficients of the quadric $q_1$, the entries where $\mu$ appears yield the coefficients of the quadric $q_2$, and the entries where $\nu$ appears give the coefficients of the quadric $q_3$. For example, the variable $\lambda$ appears in the (1, 1) and (2, 2) entries of matrix (\ref{detrep in 2.2}). Because the coefficient of $\lambda$ in the (1, 1) entry is 1 and the coefficient of $\lambda$ in the (2, 2) entry is $-1$, the equation of the quadric $q_1$ is given as $x^2-y^2$ (where the (1, 1) entry corresponds to $x^2$, and the (2, 2) entry corresponds to $y^2$). Here, $[x:y:z:w]$ denotes the coordinates of $\P^3$ (the blow-up of which yields a 1/2 Calabi--Yau 3-fold). The equations of the remaining two quadrics $q_2, q_3$ are determined in a similar fashion. The equations of the three quadrics are determined as follows:
\begin{eqnarray}
\label{three quadrics in 2.2}
q_1= & x^2-y^2 \\ \nonumber
q_2= & y^2-z^2 \\ \nonumber
q_3= & w^2.
\end{eqnarray}

\par We denote the curve in the base surface $\P^2$ of the 1/2 Calabi--Yau 3-fold dual to the $D_4$ singularity $p_1 =[0:0:1]$ of the quartic (\ref{quartic curve in 2.2}) by $l_1$. Here, $l_1$ is given by $c = 0$ in the base $\P^2$, where we use $[a:b:c]$ to denote the homogeneous coordinates of the base $\P^2$ of the 1/2 Calabi--Yau 3-folds. We denote by $l_2, l_3, l_4$ the three curves dual to the three intersection points, $p_2, p_3, p_4$, of the curve $\nu = 0$ with each of the three curves $\lambda=0, \mu=0$, and $\lambda=\mu$ yielding $A_1$ singularities. The discriminant of the extremal 1/2 Calabi--Yau 3-fold with $D_4A_1^3$ singularity given by blowing up the base points of the three quadrics (\ref{three quadrics in 2.2}) is then given by the following:
\begin{equation}
\Delta \sim l_1^6\cdot l_2^2\cdot l_3^2\cdot l_4^2.
\end{equation}

\subsection{$A_3^2A_1$ singularity}
\label{sec2.3}
\par We deduce the three quadrics yielding the extremal 1/2 Calabi--Yau 3-fold with an $A_3^2A_1$ singularity type. The plane quartic curve with $A_3^2A_1$ singularity is reducible into a conic and two lines tangent to it, as presented in Figure \ref{figD4sum3A1}. The equation of the quartic curve is as follows:
\begin{equation}
(\mu\nu-\lambda^2)\mu\nu=0,
\end{equation}
and the determinantal representation is reducible into two linear factors and a 2 $\times$ 2 submatrix. The determinantal representation is given as follows:
\begin{equation}
\label{detrep in 2.3}
\begin{pmatrix}
\mu & \lambda & 0 & 0 \\
\lambda & \nu & 0 & 0 \\
0 & 0& -\mu & 0 \\
0 & 0& 0 & -\nu
\end{pmatrix}.
\end{equation}
The equations of the three quadrics can be deduced from the representation (\ref{detrep in 2.3}) in a way similar to that discussed in section \ref{sec2.2}, and the three quadrics are given as follows:
\begin{eqnarray}
\label{three quadrics in 2.3}
q_1= & 2xy \\ \nonumber
q_2= & x^2-z^2 \\ \nonumber
q_3= & y^2-w^2.
\end{eqnarray}

\par We denote the conic $\mu\nu-\lambda^2=0$ as $C$, and use $C^{*}$ to denote the dual curve of $C$ in the base $\P^2$. We denote the curves in the base $\P^2$ of the 1/2 Calabi--Yau 3-fold dual to the two points $p_1=[0:0:1], p_2=[0:1:0]$ at which each of the two lines $\mu=0, \nu=0$ is tangent to the conic $C$ yielding $A_3$ singularities by $l_1, l_2$. $l_1$ is given by $c=0$, and $l_2$ is given by $b=0$. The curve dual to the intersection point $p_3$ of the two lines yielding an $A_1$ singularity is denoted as $l_3$. The extremal 1/2 Calabi--Yau 3-fold with an $A_3^2A_1$ singularity type then has the following discriminant:
\begin{equation}
\Delta \sim C^{*}\cdot l_1^4\cdot l_2^4 \cdot l_3^2.
\end{equation}

\subsection{$A_7$ singularity}
\label{sec2.4}
We deduce the equations of the three quadrics yielding the extremal 1/2 Calabi--Yau 3-folds with $A_7$ singularity. The quartic curve in $\P^2$ possessing $A_7$ singularity is two conics meeting at one point, as presented in Figure \ref{figA7}. The equation of this quartic curve is given as follows:
\begin{equation}
\label{quartic curve in 2.4}
(\lambda\nu-\mu^2)\big(\lambda(\lambda+\nu)-\mu^2\big)=0.
\end{equation}
The determinantal representation of the quartic curve is given in the following:
\begin{equation}
\label{detrep in 2.4}
\begin{pmatrix}
\lambda & \mu & 0 & 0 \\
\mu & \nu & 0 & 0 \\
0 & 0& \lambda & \mu \\
0 & 0& \mu & \lambda+\nu
\end{pmatrix}.
\end{equation}
The equations of the three quadrics can be deduced from the representation (\ref{detrep in 2.4}), and the three quadrics are given as follows:
\begin{eqnarray}
q_1= & x^2+z^2+w^2 \\ \nonumber
q_2= & 2xy+2zw \\ \nonumber
q_3= & y^2+w^2.
\end{eqnarray}

\par We denote the two conics of the quartic curve (\ref{quartic curve in 2.4}) by $C_1$ and $C_2$. The curve in the base $\P^2$ of the 1/2 Calabi--Yau 3-fold dual to the intersection point $p = [0:0:1]$ of the two conics $C_1$ and $C_2$ is given by $c = 0$. We denote this dual curve in the base $\P^2$ by $l$, and denote the duals of the conics $C_1$ and $C_2$ by $C_1^{*}$ and $C_2^{*}$, respectively. The discriminant of the extremal 1/2 Calabi--Yau 3-fold with an $A_7$ singularity type is then given as follows:
\begin{equation}
\Delta \sim l^8\cdot C_1^{*}\cdot C_2^{*}.
\end{equation}

\subsection{$A_5A_2$ singularity}
\label{sec2.5}
We determine the equations of the three quadrics yielding the extremal 1/2 Calabi--Yau 3-folds with $A_5A_2$ singularity. The plane quartic curve possessing $A_5A_2$ singularity is a cubic curve with a cusp and a line tangent to the flex, as presented in Figure \ref{figA7}. The equation of this quartic curve is given as follows:
\begin{equation}
\label{quartic curve in 2.5}
(\lambda^3+\mu\nu^2)\mu=0.
\end{equation}
The determinantal representation of the quartic curve is given \footnote{A discussion of a determinantal representation of a cuspidal cubic can also be found in \cite{Piontkowski}.} in the following:
\begin{equation}
\label{detrep in 2.5}
\begin{pmatrix}
-\mu & 0 & \lambda & 0 \\
0 & -\lambda & \nu & 0 \\
\lambda & \nu& 0 & 0 \\
0 & 0& 0 & \mu
\end{pmatrix}.
\end{equation}
The equations of the three quadrics are obtained from the representation (\ref{detrep in 2.5}), and the three quadrics are given as follows:
\begin{eqnarray}
q_1= & -y^2+2xz\\ \nonumber
q_2= & -x^2+w^2 \\ \nonumber
q_3= & 2yz.
\end{eqnarray}

\par The curve in the base $\P^2$ of the 1/2 Calabi--Yau 3-fold dual to the cusp $p_0=[0:1:0]$ of the cuspidal cubic in (\ref{quartic curve in 2.5}) yielding $A_2$ singularity is given by $b=0$, and we denote this curve in the base $\P^2$ by $l_0$. The cuspidal cubic $\lambda^3+\mu\nu^2=0$ in (\ref{quartic curve in 2.5}) is denoted as $B$, and the dual curve in the base $\P^2$ is denoted as $B^{*}$. The curve in the base $\P^2$ dual to the flex $p_1 = [0:0:1]$ in the cuspidal cubic $B$ yielding an $A_5$ singularity is given by $c = 0$ in the base; we denote this curve in the base $\P^2$ as $l_1$. The discriminant of the extremal 1/2 Calabi--Yau 3-fold is then given by the following:
\begin{equation}
\Delta \sim l_0^3\cdot l_1^6\cdot B^{*}.
\end{equation}

\subsection{$D_6A_1$ singularity}
\label{sec2.6}
We determine the equations of the three quadrics yielding extremal 1/2 Calabi--Yau 3-folds with $D_6A_1$ singularity. The quartic curve in $\P^2$ possessing $D_6A_1$ singularity is a conic and a tangent line to it, and another line passing through the tangent point, as presented in Figure \ref{figD6A1}. The equation of this quartic curve is given as follows:
\begin{equation}
\label{quartic curve in 2.6}
(\lambda\nu-\mu^2)\lambda\mu=0.
\end{equation}
The determinantal representation of the quartic curve is given in the following:
\begin{equation}
\label{detrep in 2.6}
\begin{pmatrix}
\lambda & \mu & 0 & 0 \\
\mu & \nu & 0 & 0 \\
0 & 0 & \lambda & 0 \\
0 & 0& 0 & \mu
\end{pmatrix}.
\end{equation}
The equations of the three quadrics are deduced from the representation (\ref{detrep in 2.6}), and the three quadrics are given as follows:
\begin{eqnarray}
q_1= & x^2+z^2\\ \nonumber
q_2= & w^2+2xy \\ \nonumber
q_3= & y^2.
\end{eqnarray}

\par The conic $\lambda\nu-\mu^2=0$ in the quartic (\ref{quartic curve in 2.6}) is denoted as $C$. We denote by $l_1$ the curve in the base $\P^2$ of the 1/2 Calabi--Yau 3-fold dual to point $p_1$ at which the line $\lambda = 0$ is tangent to the conic $C$ yielding $D_6$ singularity. We denote by $l_2$ the curve in the base $\P^2$ dual to the intersection $p_2$ of the line $\mu = 0$ and the conic $C$ yielding $A_1$ singularity. The discriminant of the extremal 1/2 Calabi--Yau 3-fold with $D_6A_1$ singularity type is then given by the following:
\begin{equation}
\Delta \sim l_1^8\cdot l_2^2\cdot C^{*}.
\end{equation}

\subsection{$E_7$ singularity}
\label{sec2.7}
We determine the equations of the three quadrics yielding the extremal 1/2 Calabi--Yau 3-folds with the $E_7$ singularity type. The quartic curve in $\P^2$ possessing the $E_7$ singularity type is a cubic with a cusp and the cuspidal tangent, as presented in Figure \ref{figD6A1}. The equation of this quartic curve is given as follows:
\begin{equation}
\label{quartic curve in 2.7}
(\lambda^3+\mu\nu^2)\nu=0.
\end{equation}
The determinantal representation of the quartic curve is given in the following:
\begin{equation}
\label{detrep in 2.7}
\begin{pmatrix}
-\mu & 0 & \lambda & 0 \\
0 & -\lambda & \nu & 0 \\
\lambda & \nu& 0 & 0 \\
0 & 0& 0 & \nu
\end{pmatrix}.
\end{equation}
The equations of the three quadrics are deduced from the representation (\ref{detrep in 2.7}), and the three quadrics are given as follows:
\begin{eqnarray}
q_1= & -y^2+2xz \\ \nonumber
q_2= & -x^2 \\ \nonumber
q_3= & 2yz+w^2.
\end{eqnarray}

\par The cuspidal cubic $\lambda^3+\mu\nu^2=0$ in (\ref{quartic curve in 2.7}) is denoted as $B$. The curve in the base surface $\P^2$ of the 1/2 Calabi--Yau 3-fold dual to the $E_7$ singularity at the cusp $p=[0:1:0]$ of the cuspidal cubic $B$ is denoted as $l$; in addition, the curve $l$ is given by $b = 0$. The discriminant of the extremal 1/2 Calabi--Yau 3-fold with singularity type $E_7$ is then given as follows:
\begin{equation}
\Delta \sim l^9\cdot B^{*}.
\end{equation}

\section{Singular fibers of extremal 1/2 Calabi--Yau 3-fold}
\label{sec3}
Studying the equations of the three quadrics of the extremal 1/2 Calabi--Yau 3-folds, we analyze the singular fibers. We perform operations of blow-ups to conduct this analysis. The extremal 1/2 Calabi--Yau 3-folds with the singularity types $D_4A_1^3$ and $A_3^2A_1$ require only shallow levels of blow-ups to understand the structures of the singular fibers and sections. The results of the blow-ups of the two singularity types are described in sections \ref{sec3.1} and \ref{sec3.2}. While the remaining four types of singularities require deeper levels of blow-ups, the singular fibers and sections are expected to be understood in a similar fashion. We discuss these cases in section \ref{sec3.3}. 

\subsection{Extremal 1/2 Calabi--Yau 3-fold with $D_4A_1^3$ singularity}
\label{sec3.1}
\par We study the three quadrics yielding the 1/2 Calabi--Yau 3-fold with the $D_4A_1^3$ singularity deduced in section \ref{sec2.2} to demonstrate that this extremal 1/2 Calabi--Yau 3-fold has type $I_0^*$ fibers. 
\par From the equations of the three quadrics (\ref{three quadrics in 2.2}), we can see that there are four base points: $[1:\pm1:\pm 1:0]$. Because there are generally eight base points given three quadrics, this result implies that each of the four base points has multiplicity 2. This can also be viewed as four points and an additional four points `` infinitely near'' to them. 
\par
\begin{figure}
\begin{center}
\includegraphics[height=10cm, bb=0 0 960 540]{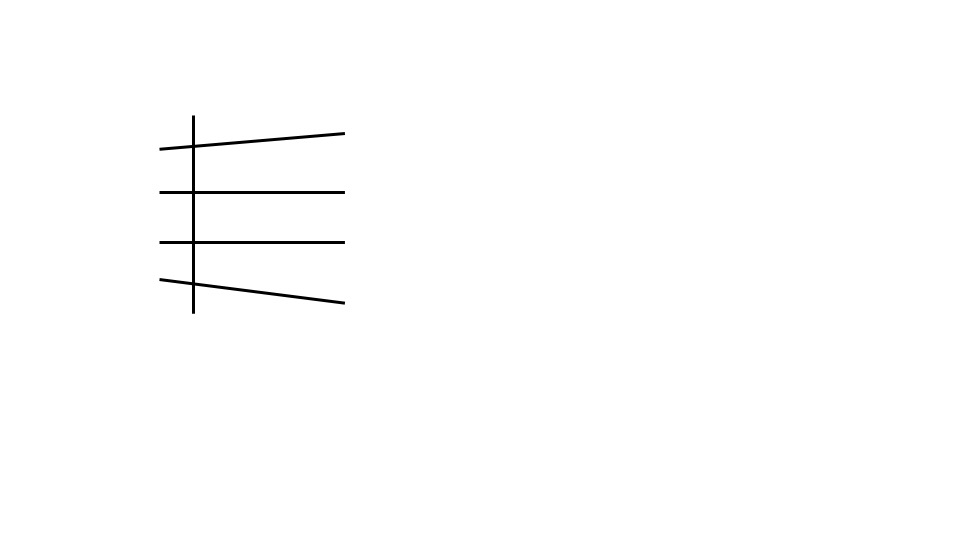}
\caption{\label{figI0}The image shows a singular fiber corresponding to $D_4$ singularity. The vertical line represents a conic, and the four horizontal lines represent $\P^1$s that result from the initial four blow-ups. This describes a type $I_0^*$ fiber.}
\end{center}
\end{figure}

 Blowing up the four base points $[1:\pm1:\pm 1:0]$ separates the four points and the four points infinitely near to them. From the equations of the three quadrics, we can see that the singular fiber corresponding to the $D_4$ singularity is 
\begin{eqnarray}
\label{double conics in 3.1}
b(x^2-y^2)-a(y^2-z^2)= & 0 \\ \nonumber
w^2= & 0,
\end{eqnarray}
where $a,b$ denote parameters such that $[a:b]$ parameterizes the discriminant component. The equation (\ref{double conics in 3.1}) represents a double conic. Because the double conic (\ref{double conics in 3.1}) contains the four base points, when the four base points are blown up, four $\P^1$s arise from the four base points in the double conic. As a result of the four blow-ups, the singular fiber corresponding to $D_4$ singularity is described as a conic and four $\P^1$s, each of which intersects with the conic in one point. Therefore, we can explicitly see that, after the four blow-ups, type $I_0^*$ fibers appear. This situation is shown in Figure \ref{figI0}. 

\par When the total space $\P^3$ is considered, after these four blow-ups occur at the four base points, four $\P^2$s arise. Each of the $\P^2$s contains one point of indeterminacy. The morphism from each $\P^2$ arising from the blow-ups to the base $\P^2$ is not a surjection, and the image is isomorphic to $\P^1$. We consider the blow-ups of the four points of indeterminacy. These additional four blow-ups transform the four $\P^2$s that appeared from the previous four blow-ups into Hirzebruch surface $\mathbb{F}_1$s, and $\P^2$ arises from each of the four $\mathbb{F}_1$s. Each of the four $\P^2$s that arose from the latter four blow-ups surjects onto the base $\P^2$ under the projection. They yield sections and generate the Mordell--Weil group.

\subsection{Extremal 1/2 Calabi--Yau 3-fold with $A_3^2A_1$ singularity}
\label{sec3.2}
Next, we analyze the singular fibers of the extremal 1/2 Calabi--Yau 3-fold with the $A_3^2A_1$ singularity type by conducting a blow-up. The base points of the three quadrics (\ref{three quadrics in 2.3}) consist of four points: $[0:1:0:\pm 1]$, $[1:0:\pm 1:0]$. The singular fibers corresponding to one of the two $A_3$ singularities are given by the following:
\begin{eqnarray}
\label{two conics in 3.2}
2b\, xy-a(x^2-z^2)= & 0 \\ \nonumber
y^2-w^2= & 0,
\end{eqnarray}
($[a:b]$ parameterizes the discriminant component.) The result of the singular fibers corresponding to the other $A_3$ singularity is analogous to what we now describe. Without a blow-up, (because $y^2-w^2=0$ in (\ref{two conics in 3.2}) splits into two linear factors,) one can only find two conics intersecting at two points, and the type $I_4$ fiber is not apparent. 
\par
\begin{figure}
\begin{center}
\includegraphics[height=10cm, bb=0 0 960 540]{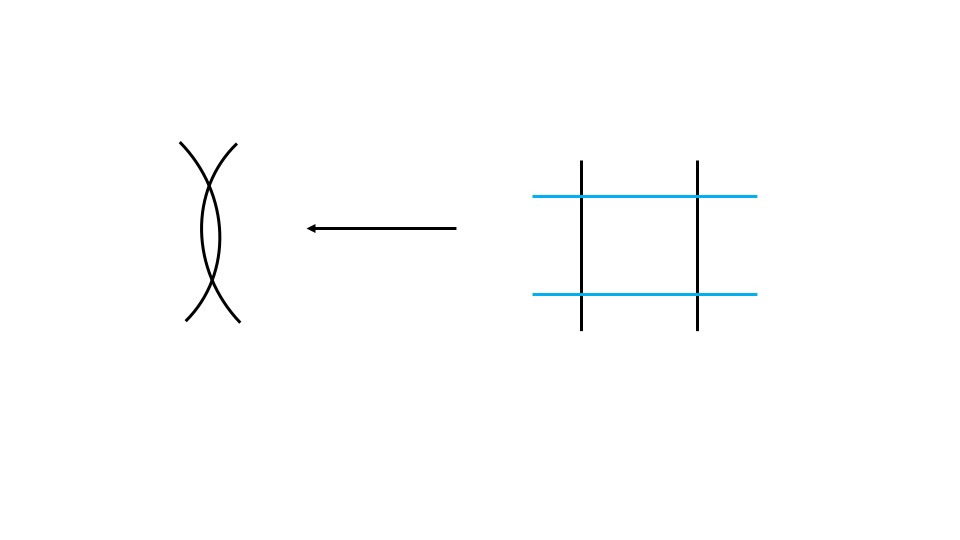}
\caption{\label{figI4}The left image represents two conics meeting at two points before the blow-ups. The right image describes type $I_4$ fiber after two blow-ups. The two vertical lines are two separated conics, and the horizontal blue lines are two $\P^1$s that arise after two blow-ups.}
\end{center}
\end{figure}

 The two conics in (\ref{two conics in 3.2}) intersect at two points $[1:0:\pm 1:0]$, which are two among the base points. When these points are blown up, two intersecting conics are separated as a result of the two blow-ups, and two $\P^1$s arise from the two intersection points. Consequently, the structure of the type $I_4$ fibers becomes clear after the two blow-ups, as described in Figure \ref{figI4}. When the remaining two base points $[0:1:0: \pm 1]$ are blown up, the structure of the type $I_4$ fibers can also be explicitly seen from the other $A_3$ singularity. 

\par When conducting four blow-ups of the four points, four $\P^2$s arise from the total space $\P^3$. When four additional blow-ups of the resulting $\P^2$s are applied, these latter four blow-ups transform the four $\P^2$s that arise from the previous four blow-ups into four $\mathbb{F}_1$s, and the four $\P^2$s that arise from the latter four blow-ups surject onto $\P^2$ through projection, yielding global sections, generating the Mordell--Weil group.

\subsection{Extremal 1/2 Calabi--Yau 3-folds with other singularity types}
\label{sec3.3}
Extremal 1/2 Calabi--Yau 3-folds with the other four singularity types require deeper levels of blow-ups to analyze the singular fibers. The extremal 1/2 Calabi--Yau 3-folds with two singularity types discussed in sections \ref{sec3.1} and \ref{sec3.2} required two stages of blow-ups: the first four blow-ups and the next four blow-ups. After these blow-ups, the structures of the singular fibers and sections can clearly be seen. 
 \par For example, concerning the extremal 1/2 Calabi--Yau 3-folds with $A_7$ and $A_5A_2$ singularities before a blow-up, only two conics meeting at two points can be seen from the $A_7$ and $A_5$ singularities. The base points of the quadrics yielding extremal 1/2 Calabi--Yau 3-folds with the two singularities $A_7$ and $A_5A_2$ consist of two and three points, respectively, implying that extremal 1/2 Calabi--Yau 3-folds with $A_7$ and $A_5A_2$ singularities require deeper stages of blow-ups to analyze the singular fibers than those described in sections \ref{sec3.1} and \ref{sec3.2}. It is expected that, after multiple stages of blow-ups, $\P^1$s forming an octagon and hexagon can be seen from $A_7$ and $A_5$ singularities, yielding type $I_8$ and $I_6$ fibers. 
\par For extremal 1/2 Calabi--Yau 3-folds with $D_6A_1$ and $E_7$ singularities, after multiple stages of blow-ups, we expect that type $I_2^*$ and type $III^*$ fibers can be explicitly seen from the $D_6$ and $E_7$ singularities. A future study can focus on investigating the detailed structures of extremal 1/2 Calabi--Yau 3-folds with these singularity types.

\section{Application to 6D F-theory}
\label{sec4}
Taking double covers of 1/2 Calabi--Yau 3-folds ramified over a hypersurface of degree 4 in terms of the three quadrics $q_1, q_2, q_3$ yields elliptic Calabi--Yau 3-folds \cite{Kimura1910}. Seven tensor fields arise \footnote{The base surface of the elliptic Calabi--Yau 3-folds as double covers of the 1/2 Calabi--Yau 3-folds is isomorphic to a degree-2 del Pezzo surface \cite{Kimura1910}.} in 6D $N = 1$ F-theory on the resulting Calabi--Yau 3-folds \cite{Kimura1910}. The resulting Calabi--Yau 3-fold and the original 1/2 Calabi--Yau 3-fold have an identical singularity type \cite{Kimura1910}. Therefore, taking double covers of the extremal 1/2 Calabi--Yau 3-folds yields elliptic Calabi--Yau 3-folds with singularity types $E_7$, $D_6A_1$, $A_7$, $A_5A_2$, $A_3^2A_1$, and $D_4A_1^3$. $E_7$ gauge group forms in 6D F-theory on the Calabi--Yau 3-fold obtained as double cover of 1/2 Calabi--Yau 3-fold with $E_7$ singularity type as constructed in section \ref{sec2.7}. 
\par The sum of the Mordell--Weil rank and the rank of the singularity type of a 1/2 Calabi--Yau 3-fold is always seven \cite{Kimura1910}, and therefore every extremal 1/2 Calabi--Yau 3-fold has Mordell--Weil rank 0. A Calabi--Yau 3-fold as a double cover of an extremal 1/2 Calabi--Yau 3-fold has Mordell--Weil rank equal to or greater than the extremal 1/2 Calabi--Yau 3-fold \cite{Kimura1910}, and thus not much can be stated regarding the U(1) symmetry forming in 6D F-theory on the resulting Calabi--Yau 3-folds constructed as double covers of extremal 1/2 Calabi--Yau 3-folds.
\par The method relating the singularity types of 1/2 Calabi--Yau 3-folds to those of the quartic curves in $\P^2$ discussed in section \ref{sec2.1} also applies to 1/2 Calabi--Yau 3-folds of lower singularity ranks. Thus, we obtain the classification of the singularity types of 1/2 Calabi--Yau 3-folds from the classification results of the plane quartic curves \cite{DolgachevAlgGeom}, by applying the method in \cite{Muk, Mukai2008, Mukai2019}. Concrete constructions of the subextremal 1/2 Calabi--Yau 3-folds (namely 1/2 Calabi--Yau 3-folds possessing the singularity types of rank 6) can be a likely target of future studies. Because the subextremal 1/2 Calabi--Yau 3-folds have Mordell--Weil rank 1, their double covers yield Calabi--Yau 3-folds of Mordell--Weil rank of at least 1. 6D $N = 1$ F-theory compactifications on the resulting Calabi--Yau 3-folds have (at least) one U(1) factor. 
\par We utilized the ``duality'' of the singularities of the quartic curves in $\P^2$ and the 1/2 Calabi--Yau 3-folds to deduce the equations of the three quadrics yielding the extremal 1/2 Calabi--Yau 3-folds. The discriminant of a Calabi--Yau 3-fold constructed as a double cover of a 1/2 Calabi--Yau 3-fold can be deduced from the discriminant of the 1/2 Calabi--Yau 3-fold \cite{Kimura1910}. The discriminants of the extremal 1/2 Calabi--Yau 3-folds were deduced in sections \ref{sec2.2} - \ref{sec2.7} by utilizing the duality of the singularity types of the plane quartic curves and the 1/2 Calabi--Yau 3-folds. This method also applies to 1/2 Calabi--Yau 3-folds of lower singularity ranks; therefore, the discriminants of the elliptic Calabi--Yau 3-folds as their double covers can also be deduced in a similar manner. Because matter fields localize at the intersections of the 7-branes wrapped on the discriminant components, the locations of the localized matter in 6D F-theory on the Calabi--Yau 3-folds are determined from the deduced discriminants. The base change lifts the global sections of the 1/2 Calabi--Yau 3-folds to sections of the Calabi--Yau 3-fold as their double covers \cite{Kimura1910}. 
\par Because the structures of the singular fibers can be analyzed through blow-up operations as demonstrated in sections \ref{sec3.1} and \ref{sec3.2}, there is a chance that the matter spectra in 6D F-theory on the Calabi--Yau 3-folds can also be deduced by studying the structures of the singular fibers at the collisions of the fibers, which correspond to the intersections of the 7-branes, an investigation into which can be a direction of future study. Because the blow-up methods described in sections \ref{sec3.1} and \ref{sec3.2} revealed the structures of the sections, they might be used to analyze the explicit forms of the sections. The data mentioned can be used to determine the charges of the hypermultiplets charged under the $U(1)$ gauge symmetries \cite{Kimura1910}, when 6D F-theory is compactified on Calabi--Yau 3-folds of positive Mordell--Weil ranks \footnote{The Mordell--Weil ranks of Calabi--Yau 3-folds constructed as the double covers of 1/2 Calabi--Yau 3-folds of positive Mordell--Weil ranks are positive \cite{Kimura1910}.} constructed as double covers of 1/2 Calabi--Yau 3-folds.

\section{Open problems}
\label{sec5}
We determined the singularity types of extremal 1/2 Calabi--Yau 3-folds, i.e., 1/2 Calabi--Yau 3-folds with a singularity of rank 7, and deduced the equations of the three quadrics yielding the extremal 1/2 Calabi--Yau 3-folds. Double covers of the extremal 1/2 Calabi--Yau 3-folds give elliptic Calabi--Yau 3-folds, the singularity types of which are identical to those of the original 1/2 Calabi--Yau 3-folds \cite{Kimura1910}. The methods we described in sections \ref{sec3.1} and \ref{sec3.2} enabled us to analyze the geometric structures of the singular fibers and sections. These methods might also have applications in investigating the matter fields arising in 6D F-theory on the Calabi--Yau 3-folds as double covers. 
\par Seven tensor fields arise in 6D $N = 1$ F-theory on elliptic Calabi--Yau 3-folds as double covers of 1/2 Calabi--Yau 3-folds \cite{Kimura1910}. The analysis in \cite{Kimura1910} suggests that 6D F-theory models with $T = 7$ on Calabi--Yau 3-folds as double covers of 1/2 Calabi--Yau 3-folds are contained in vast numbers of models of 6D F-theory with $T = 7$, based on the fact that non-Abelian gauge groups of ranks of up to only 7 can form in 6D F-theory on Calabi--Yau 3-folds as double covers of 1/2 Calabi--Yau 3-folds. Analogous to the fact that the points in the moduli of elliptic K3 surfaces where a K3 surface splits into a pair of rational elliptic surfaces \footnote{Structure of the singular fibers when an elliptic K3 surface splits into a pair of rational elliptic surfaces were analyzed in the context of F-theory using the quadratic base change in \cite{KRES}.} correspond to the stable degeneration limit \cite{FMW, AM} at which F-theory/heterotic duality \cite{Vaf, MV1, MV2, Sen, FMW} is strictly formulated, do the points in the complex structure moduli of Calabi--Yau 3-folds where a Calabi--Yau 3-fold splits into 1/2 Calabi--Yau 3-folds correspond to certain limits with a physical significance? 
\par Furthermore, do elliptic Calabi--Yau 3-folds 6D F-theory compactifications on which have a number of tensor fields other than seven exhibit an analogous geometric structure wherein Calabi--Yau 3-folds allow splitting into building blocks of elliptic 3-folds? If so, investigating such elliptic Calabi--Yau 3-folds can be an interesting approach. If such Calabi--Yau 3-folds do not exist, do the 6D $N = 1$ models with seven tensor fields have a special meaning? Investigating these questions might be interesting in relation to the swampland conditions.

\section*{Acknowledgments}

We would like to thank Shigeru Mukai for discussions.

\end{document}